\documentclass[10pt]{article}
\usepackage{amsmath,amssymb,latexsym,amsthm,amscd,amsxtra}
\usepackage[all]{xy}
\usepackage{epsfig}
\usepackage{graphicx}   
\usepackage{longtable}
\usepackage[bottom]{footmisc}
\usepackage[section]{placeins}
\usepackage{dsfont}
\usepackage{verbatim}
\newcommand{\dd}{\! \mathrm{d}}
\newcommand{\Hhat}{\hat{H}}
\newcommand{\vbar}{|\,}
\newcommand{\psq}{\hat{p}\,^2}

\begin{document}

\begin{center}
{\bf \Large Sharp and Infinite Boundaries in the Path Integral Formalism}

\vspace*{0.25in}
{\large Phillip Dluhy\footnote{pdluhy@luc.edu} and Asim Gangopadhyaya\footnote{agangop@luc.edu}}

\vspace*{0.05in}
{Loyola University Chicago, Department of Physics, Chicago,
IL 60660}
\end{center}
\date{\today}


\begin{abstract}
We revisit the analysis of sharp infinite potentials within the path integral formalism using the image method \cite{Goodman}. We show that the use of a complete set of energy eigenstates that satisfy the boundary conditions of an infinite wall precisely generates the propagator proposed in Ref. \cite{Goodman}. We then show the validity of the image method by using supersymmetric quantum mechanics to relate a potential without a sharp boundary to the infinite square well and derive its propagator with an infinite number of image charges. Finally, we show that the image method readily generates the propagator for the half-harmonic oscillator, a potential that has a sharp infinite boundary at the origin and a quadratic potential in the allowed region, and leads to the well known eigenvalues and eigenfunctions.
\end{abstract}

\vspace*{0.25in}

\noindent
{\bf \Large Introduction}
\vspace*{0.05in}

\noindent
Sharp and infinite potentials are difficult to handle in the path integral formalism \cite{Feynman,Kleinert,Schulman}. Goodman \cite{Goodman}, in one of the first papers to address this problem, advocated the use of the image-point  method to account for paths that wander into the forbidden region. He applied this intuitive method to the infinite square well and generated the correct set of eigenvalues and eigenfunctions. In Ref. \cite{Sokmen-ISW}, S\"okmen provided additional support for the image method by connecting the infinite square well (ISW) to the Rosen-Morse potential via a change of variables (point canonical transformation). Starting with the Rosen-Morse potential, which has no sharp boundary, S\"okmen derived the propagator of Ref. \cite{Goodman}, and thus verified, at least for the case of infinite square well, the validity of the image method. Auerbach and Kivelson \cite{Auerbach_Kivelson}, Nevels, Wu and Huang \cite{Nevels_Wu_Huang}, and Auerbach and Schulman \cite{Auerbach_Schulman} have provided much more rigorous reasoning to support the image method and have extended it to systems with non-zero potentials in the allowed region .
\begin{figure}[htb]
\centering
\includegraphics[width=4in]{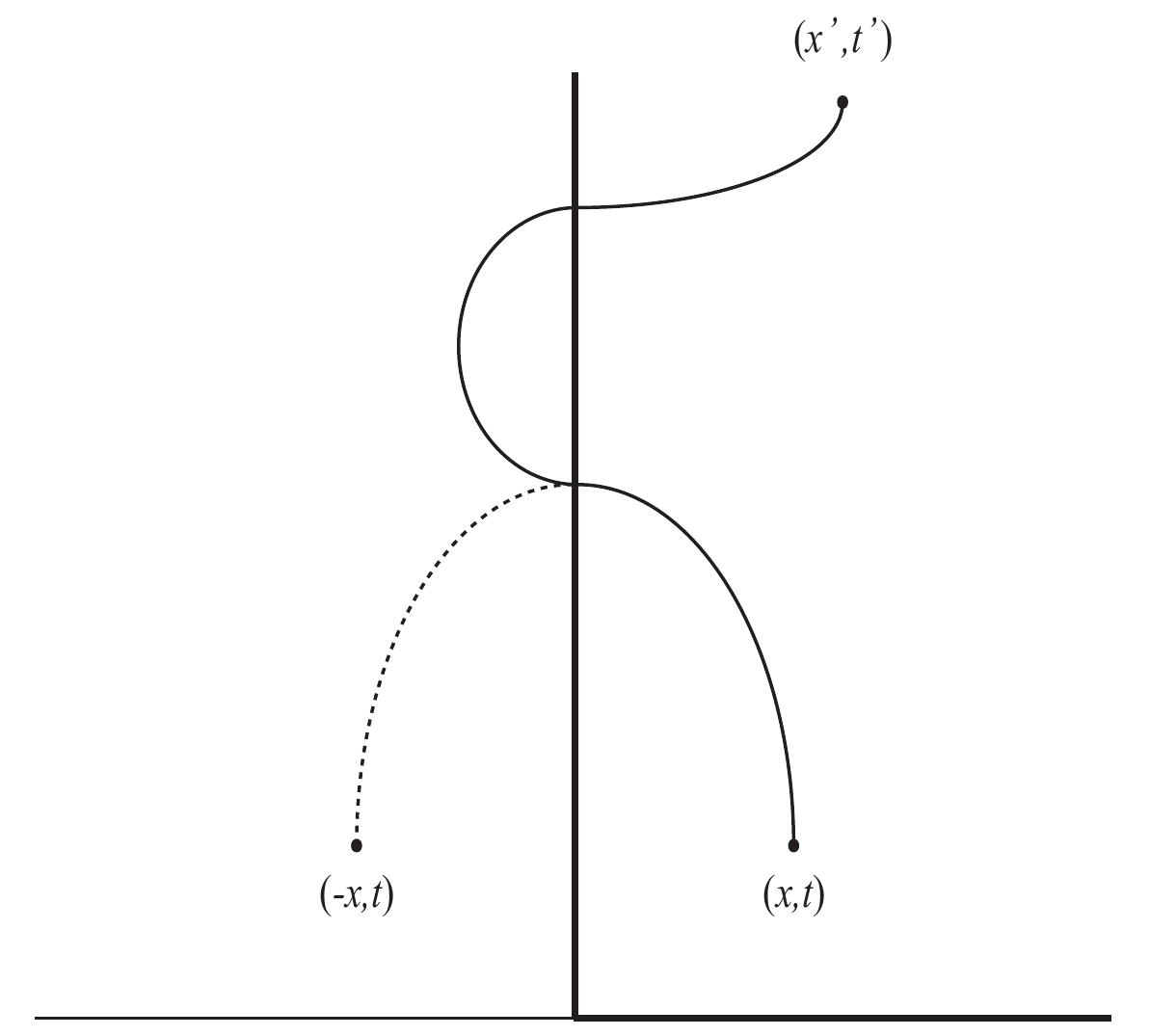}\\
\caption{Inclusion of the term originating from the image-point ($-x,t$) has the effect of excluding all forbidden paths such as the one shown by the solid line.}\label{Diagram1}
\end{figure}

In this paper, we aim to provide further support for the validity of Goodman's conjecture. First, we show that the use of eigenfunctions that vanish at the infinite boundary automatically generates the propagator Goodman found for a particle in a restricted domain, i.e. it naturally generates the propagator with an image point. We then, as S\"okmen did in Ref. \cite{Sokmen-ISW}, generate the correct propagator for the ISW that contains an infinite number of image charges, starting from the Rosen-Morse potential.  However, instead of using a point canonical transformation \cite{Gervais,Dutt2,Dutt3} to map the Rosen-Morse potential to the ISW, we show that a simple change of parameter is sufficient for this mapping. We use supersymmetric quantum mechanics to show the validity of this limiting procedure. Finally, we use the image method to derive the propagator for a half-harmonic oscillator with a sharp boundary at the origin and show that it generates the well known spectrum for this system. This final case provides an explicit example of a particle in a non-zero potential with a sharp boundary.

\vspace*{0.25in}
{\bf \Large The propagator for a free particle in a restricted domain}

In this section we will derive the propagator for a particle in the presence of an infinite potential barrier using the path integral approach. The particle is free in the allowed region: $0< x<\infty$.  The Hamiltonian for this system can be written as $\Hhat = \frac{\psq}{2m}+V(x)$, where $V(x)$ is given by
\begin{eqnarray}
V(x)=\left\{
\begin{array}{l l}
\infty &\quad \text{for $x \le0$}\\
0 &\quad \text{for $x > 0$}\\
\end{array}\right.~.
\end{eqnarray}
Our goal is to determine the propagator $K_R(x', t'; x,t) \equiv \, _{\cal H}\!\langle x', t'\vbar\, x,t\rangle_{\cal H} \equiv \langle x'\vbar e^{-\frac{i}{\hbar}\Hhat (t'-t)} \vbar x\rangle$ for this system that respects its boundary conditions. We have used the notation $\vbar x, t\rangle_{\cal H}$ to denote an eigenstate of the position operator in the Heisenberg representation that is related to an eigenstate in the Schr\"odinger picture by $ \vbar x, t\rangle_{\cal H}= e^{i \Hhat t/\hbar}\vbar x \rangle$. To simplify the notation, we will drop the subscript $\cal H$.

Following Ref. \cite{Abers_Lee}, we split the time-interval $(t'-t)$ into $n+1$ segments, each of size $\epsilon$, letting $t_{\!\ell}$ denote the time $t+\ell\epsilon$, with $t+(n+1)\,\epsilon = t'$.
Now, using the completeness of state vectors $\vbar x_{\!\ell}, t_{\!\ell} \rangle$, i.e., $\int \dd x_{\!\ell}~ \vbar x_{\!\ell} ,t_{\!\ell} \rangle\langle  x_{\!\ell} ,t_{\!\ell} |=\mathds{1},$ where  $x_{\!\ell}$ denotes position at an intermediate time $t_{\!\ell}$, we can write the matrix element as
\begin{eqnarray}
\langle x', t'\vbar x,t\rangle =
\int \dd x_{n}\int \dd x_{n-1} \int \! \cdots
 \int \dd x_{1}  \, \langle x', t'\vbar x_n,t_n\rangle \cdots  \langle x_{\!\ell}, t_{\!\ell}\vbar x_{\ell-1},t_{\ell-1} \rangle\cdots \langle x_1,t_1\vbar x,t \rangle~.
\end{eqnarray}
For very large $n$, and hence for an infinitesimally small $\epsilon$,  the propagator
$\langle x_{\!\ell}, t_{\!\ell}\vbar x_{\ell-1},t_{\ell-1} \rangle$ can be written as
\begin{eqnarray}
\langle x_{\!\ell}, t_{\!\ell}\vbar x_{\ell-1}, t_{\ell-1} \rangle &\!\!\!\!=&\!\!\!\! \left\langle x_{\!\ell} \left| e^{-i\frac{ \Hhat \epsilon}{\hbar}} \right| x_{\ell-1} \right\rangle = \int_0^\infty \dd E~\langle x_{\!\ell} \,| e^{-i\frac{ \Hhat \epsilon}{\hbar}}|   E\rangle \, \langle E \vbar x_{\ell-1} \rangle\nonumber\\ \nonumber \\
&& \!\!\!\!\!\!\!\! \!\!\!\!\!\!\!\! \!\!\!\!\!\!\!\! \!\!\!\!\!\!\!\! =
\int_0^\infty \dd E~\langle x_{\!\ell} \,|   E\rangle \, \langle E \vbar x_{\ell-1} \rangle
~e^{-i\epsilon\,\frac{H(p_{\!\ell})}{\hbar}}
=
\int_0^\infty \frac{2~\dd k_{\!\ell}}{\pi}~\sin k_{\!\ell} x_{\!\ell}~ \sin k_{\!\ell} x_{\ell-1}
~e^{-i\epsilon\,\frac{H(p_{\!\ell})}{\hbar}} \label{Infintesimal-matrix}~,
\end{eqnarray}
where the function $H(p_{\!\ell})$, without a hat, represents the classical Hamiltonian $\frac{p_{\!\ell}^2}{2m}$.
The kets $| E\rangle$ represent the energy eigenstates for a free particle in the domain $0<x<\infty$. The corresponding coordinate representation is given by 
$\langle x \vbar E\rangle = \sqrt{\frac{2m}{\pi\hbar^2 k}} ~\sin kx,$
which vanishes at the infinite wall at $x=0$. The wavenumber $k$ is related to the energy by  $E=\frac{\hbar^2\, k^2}{2m}$. The factor in front of $\sin kx$ is derived from the requirement that $\int \dd E \langle x \vbar E\rangle \langle E\vbar y \rangle$ must equal $\delta(x-y)$, the completeness condition. To compute the above infinitesimal propagator $\langle x_{\!\ell}, t_{\!\ell}\vbar x_{\ell-1},t_{\ell-1} \rangle$, following Ref. \cite {Nevels_Wu_Huang}, we write $\int_0^\infty \frac{2~\dd k_{\!\ell}}{\pi}~\sin k_{\!\ell} x_{\!\ell}~ \sin k_{\!\ell} x_{\ell-1}$ as
\begin{eqnarray}
 \int_0^\infty \frac{2~\dd k_{\!\ell} }{\pi}\,&&
 \! \! \! \! \! \! \! \! \! \! \! \! \!
 \sin \left(k_{\!\ell}\, x_{\!\ell}\right) \, \sin \left(k_{\ell }\,x_{\ell -1}\right)
 =   -\frac12 \int_0^\infty \frac{\dd k_{\!\ell} }{\pi}~\left[e^{ik_{\!\ell} x_{\!\ell} }-e^{-ik_{\!\ell} x_{\!\ell} }\right]~\left[e^{ik_{\ell}x_{\ell -1}}-e^{-ik_{\ell}x_{\ell -1}}\right]
 \nonumber\\
 \nonumber\\ &\!\!\!\!=&\!\!\!\!
  \int_{-\infty}^\infty \frac{\dd k_{\!\ell} }{2\pi}~ \left[e^{ik_{\!\ell} (x_{\!\ell} -x_{\ell -1})}-e^{ik_{\!\ell} (x_{\!\ell} +x_{\ell -1})}\right]~. \label{Sine-Exponential}
\end{eqnarray}
Substituting\footnote{Such a substitution is justified because the transformation $k\rightarrow\, -k$ has no effect on the Hamiltonian as it is a quadratic function of the momentum.} this expression into Eq. (\ref{Infintesimal-matrix})
yields\begin{eqnarray}
\langle x_{\!\ell}, t_{\!\ell}\,\vbar x_{\ell-1}, t_{\ell-1} \rangle
=
  \int_{-\infty}^\infty \frac{\dd k_{\ell}}{2\pi}~ \left[e^{ik_{\!\ell} (x_{\ell}-x_{\!\ell-1})}-e^{ik_{\!\ell} (x_{\ell}+x_{\!\ell-1})}\right]~e^{-i\epsilon\,\frac{H(p_{\ell})}{\hbar}}~.
\label{Infintesimal-matrix1}
\end{eqnarray}
Inserting this infinitesimal propagator into Eq. (2), we get
\begin{align*}
 \langle x', t'\vbar  x,t\rangle =& \prod_{\ell=1}^n \int_0^\infty\dd x_{\!\ell} \prod_{\ell=1}^{n+1}\left[\int_{-\infty}^{\infty} \!\frac{\mathrm{d}k_{\!\ell}}{2\pi}~ \left(e^{ik_{\!\ell}(x_{\ell}-x_{\!\ell-1})}-e^{ik_{\!\ell}(x_{\ell}+x_{\!\ell-1})}\right)~e^{-i\hbar\epsilon k_{\!\ell}^2/2m}\right]~.
\end{align*}
 These Gaussian integrals can be carried out using the identity $\int_{-\infty}^{\infty} \dd x\,  e^{i\alpha x^2} =  \sqrt{\frac{\pi i}{\alpha}}$. They yield
\begin{align*}
 \langle x', t'\vbar  x,t\rangle =& \prod_{\ell=1}^n \int_0^\infty\dd x_{\!\ell} \prod_{\ell=1}^{n+1}\left(\frac{m}{2i\epsilon\pi\hbar}\right)^{1/2}\left[e^{im(x_{\!\ell}-x_{\ell-1})^2/2\hbar\epsilon}-e^{im(x_{\!\ell}+x_{\ell-1})^2/2\hbar\epsilon}\right]\\
\end{align*}
Terms in the square bracket, upon integration, lead to free-particle propagators from
$(x,t) \rightarrow (x', t')$ and $(-x,t) \rightarrow (x', t')$ respectively\footnote{See pages 31-37 of Ref. \cite{Schulman}.}.  I.e.
\begin{align*}
 \langle x', t'\vbar  x,t\rangle =& \left(\frac{m}{2i\pi\hbar(t'-t)}\right)^{1/2} \left[e^{im(x'-x)^2/2\hbar (t'-t)}-e^{im(x'+x)^2/2\hbar (t'-t)}\right]\\
\nonumber\\
=& \langle x',t'\vbar x,t\rangle_F - \langle x',t'| -x,t \rangle_F = K_F(x',t';x,t) - K_F(x',t';-x,t)\; ,
\end{align*}
where the subscript $F$ denotes a propagator for a particle that is free in the region $-\infty < x < \infty$. Thus, we find that the propagator $K_R(x', t'; x,t)$ in the restricted domain $0 < x < \infty$ is given by the difference of free particle propagators originating from the point $(x,t)$ and its image point $(-x,t)$, as suggested in Ref. \cite{Goodman}.

\vspace*{0.25in}
{\bf \Large Green's function for the ISW from the Rosen-Morse potential}

Supersymmetric quantum mechanics (SUSYQM) \cite{Witten,Cooper1,Junker,Cooper-Khare-Sukhatme,Gangopadhyaya-Mallow-Rasinariu} is a generalization of the ladder operator formalism for the harmonic oscillator. It makes use of first order differential operators
$A^{\pm} \equiv \left(\mp\,   \frac{d } {dx} +W(x,a)\right)$, where the superpotential $W(x,a)$ is function of $x$ and a parameter $a$, and  embodies the interaction in this formalism (instead of the potential.)   The operators $A^{\pm}$ are hermitian conjugates of each other and help us define two partner hamiltonians
\begin{eqnarray} H_{\mp}= A^{\pm}\, A^{\mp}
= \left(\mp {d \over {dx}} +W(x,a)\right) \left(\pm  {d \over {dx}} + W(x,a) \right)
=-{{d^2} \over {dx^2}}+ W^2(x,a)\mp
{{dW(x,a)} \over{dx}}
=- {{d^2} \over {dx^2}}+ V_{\mp}(x,a)~,\nonumber
\label{Schrodinger}
\end{eqnarray}
where we have set $\hbar = 2m =1$.
A superpotential $W(x,\alpha)$ generates two Hamiltonians $H_\pm$ with potentials $V_{\pm}(x,\alpha)$ given by
$V_{\pm}(x) \equiv W^2(x,\alpha) \pm \frac{\dd\, W}{\dd x}.$
For example, the superpotential $W(x) = \frac12 \, \omega x$ generates the partner potentials $V_{\pm}(x) = \frac14\, \omega x^2 \pm \frac12 \,\omega$; both describe one-dimensional harmonic oscillators\footnote{Note that for harmonic oscillator, both partners, $V_{+}(x)$ and  $V_{-}(x)$, have the same $x$-dependence. This feature, known as shape invariance, plays an important role in the exact solvability of this system \cite{Infeld,Gendenshtein}.}.

The infinite square well is an extensively studied system in SUSYQM \cite{Dutt}. The general Rosen-Morse and P\"oschl-Teller potentials reduce to ISW for a particular value of a parameter. We will use this relation with the Rosen-Morse potential to derive the propagator for the ISW. We begin with the superpotential for the Rosen-Morse system: $W(x,b) = -\,b \cot x$. The domain of this potential is $0<x<{\pi}$, and the partner potentials $V_{\pm}(x,b)$ are given by
\begin{eqnarray}
V_{\pm}(x,b) \equiv W^2(x,b) \pm \frac{\dd\,}{\,\,\dd x} ~W(x,b)=b(b\pm 1)\, {\rm cosec}^2\,x - b^2~~. \label{RM}
\end{eqnarray}
For an arbitrary {\bf positive value of $\boldsymbol{b}$}, the supersymmetry is said to be unbroken \cite{Junker,Cooper-Khare-Sukhatme,Gangopadhyaya-Mallow-Rasinariu}, and the potential  $V_{-}(x,b)=b(b - 1)\,{\rm cosec}^2\,x - b^2$ holds a zero energy ground state with corresponding eigenenergies given by $E^{(-)}_n\! = \! (b+n)^2 - b^2$. Note, that for $b=1$, this potential reduces to $V_{-}(x,1)=-1$. Thus, $V_-(x,1)$ describes an Infinite Square Well of width $\pi$, and with its bottom at $-1$. We will show that for $\lim_{\,b\to 1}$  the Green's function for $V_{-}(x,b)$ reduces to that of an ISW with an infinite number of image charges.

In Refs. \cite{Sokmen-ISW,Sokmen-Rotator}, the authors derived the Green's function $G(x_f, x_i, E)$ for the potential $V_{\rm PS}(x,s) = \left(s^2-\frac14\right) \, {\rm cosec}^2\,x.$ It was given by
\begin{eqnarray}
G(x_f, x_i, E) =\sum_{n=0}^\infty \frac{(n+s+\frac12)}{E-\left(n+s+\frac12\right)^2}
\frac{\Gamma (n+2s+1)}{\Gamma(n+1)} 
\left( \sin x_f \right)^{\frac12}~P^{-s}_{n+s}\left( \cos x_f \right)
\left( \left( \sin x_i \right)^{\frac12}~P^{-s}_{n+s}\left( \cos x_i \right)\right)^*,\nonumber
  \label{GPS}
\end{eqnarray}

\noindent
where $P^{\mu}_{\nu}$ represents the Associated Legendre Polynomials. For $s=b-\frac12$, the potential $V_{\rm PS}(x,s)$ and $V_{-}(x,b)$ differ only by the constant $-b^2$, and hence, they both must have identical Green's functions, the only difference being that the eigenvalues will differ by that same constant. For $b=1$, i.e. $s=\frac12$, the Green's function is given by
\begin{eqnarray}
G(x_f, x_i, E)  =\sum_{n=0}^\infty \frac{(n+1)}{E-\left(n+1\right)^2}~
\frac{\Gamma (n+2)}{\Gamma(n+1)}
\left( \sin x_f \right)^{\frac12}~P^{-\frac12}_{n+\frac12}\left( \cos x_f \right)
\left(\left( \sin x_i \right)^{\frac12}~P^{-\frac12}_{n+\frac12}\left( \cos x_i \right)\right)^*~.
  \label{GPS2}
\end{eqnarray}
Now, we use the identity\footnote{Abramowitz \& Stegun - pp. 332, eq. (8.1.2).}
\begin{eqnarray}
P^\mu_\nu(z) = \frac{1}{\Gamma(1-\mu)}~ \left(\frac{z+1}{z-1}\right)^{\frac\mu 2}~
F\left( -\nu, \nu+1; 1-\mu; \frac{1-z}{2} \right)
\end{eqnarray}
to get
\begin{eqnarray}
P^{-\frac12}_{n+\frac12}\left( \cos x_f \right) &\!\!\!\!=&\!\!\!\! \frac{1}{\Gamma\left(\frac32\right)}~ \left(\frac{\cos x_f+1}{\cos x_f-1}\right)^{-\frac14}~
F\left( -n-\frac12, ~n+\frac32;~ \frac32; ~\sin^2 \frac{x_f}{2} \right) \nonumber\\\nonumber\\
&\!\!\!\!=&\!\!\!\! \frac1{\sqrt{\pi}} \cdot e^{i\frac{\pi}{4}} \cdot \sqrt{\frac{\sin \frac{x_f}{2}} {\cos \frac{x_f}{2}}} \cdot \frac{\sin\left[(n+1) x_f  \right]}
{(n+1)\sin \left(\frac{x_f}{2}\right)}~.\label{AbramowitzStegun-pp332}
\end{eqnarray}
In deriving above, we have used the identity\footnote{Abramowitz \& Stegun - pp. 556, eq.(15.1.15).}
$ F\left( a, 1-a;~ \frac32; ~\sin^2z \right) = \frac{\sin\left[(2a-1) z \right]}{(2a-1) \sin z}.$
Substituting the above expression for $P^{-\frac12}_{n+\frac12}\left( \cos x_f \right)$, and a similar expression for $P^{-\frac12}_{n+\frac12}\left( \cos x_i \right)$, into Eq. (\ref{GPS2}), we get
\begin{eqnarray}
G(x_f, x_i, E)  &\!\!\!\!=&\!\!\!\!  \sum_{n=0}^\infty \frac{(n+1)^2}{E-\left(n+1\right)^2}~
\frac{2 
}{(n+1)^2 \pi} ~\sin\left[(n+1) x_f  \right] ~\sin\left[(n+1) x_i  \right] \nonumber\\\nonumber\\
&\!\!\!\!=&\!\!\!\! \sum_{n=0}^\infty \frac{1}{E-\underbrace{\left(n+1\right)^2}_{E_n}}~
~\underbrace{ \sqrt{\frac{2}{\pi}}\sin\left[(n+1) x_f  \right] }_{\psi_n(x_f)}
~\underbrace{ \sqrt{\frac{2}{\pi}}\sin\left[(n+1) x_i  \right] }_{\psi_n(x_i)}~,
  \label{GPS3}
\end{eqnarray}
which clearly produces the correct spectrum for the infinite square well with ground state energy equal to $b^2 =1$.

Thus, we were able to connect the Rosen-Morse potential to the ISW simply by considering the system for a specific value of the parameter $b$. By taking the limit of $b\rightarrow 1$, we have avoided employing the point canonical transformation (PCT) of Ref. \cite{Sokmen-ISW}, which required changes of both the dependent and independent variables \footnote{It is worth pointing out that point canonical transformation often provides a way to derive information about one system by connecting it with another system \cite{Hexagon}.  A substantive research on point canonical transformation in path integral formalism was carried out in Refs. \cite{Gervais,Dutt2,Dutt3}.}. This much simpler method of mapping the Green's function of a continuous potential to the Green's function for a system with two sharp, infinite boundaries has allowed us to generate the propagator containing infinitely many image points with considerable ease.

\vspace*{0.25in}
{\bf \Large The propagator for a half-harmonic oscillator potential}

As a further verification of the image method, we will determine the propagator for a half-harmonic oscillator (H-HO) using the image method, and thence its spectrum.  In Ref. \cite{Auerbach_Schulman}, the authors  use the path decomposition expansion (PDX) \cite{Auerbach_Kivelson} to validate the image method for a general potential in the allowed region. We seek to verify their proposition by showing that the image method readily produces the propagator for the half harmonic oscillator, a potential defined over the half line $0< x<\infty$. Only the odd energy levels of the harmonic oscillator are expected to correspond to the eigenvalues and eigenfunctions of the H-HO, since the even levels would not satisfy the boundary condition that the eigenfunction must vanish at $x=0$.

The H-HO can be viewed as a system defined by
\[
V(x)=\left\{
\begin{array}{l l}
\infty &\quad \text{for $x \le0$}\\
\frac12 k x^2 &\quad \text{for $x > 0$}\\
\end{array}\right.~,
\]
containing a sharp and infinite boundary at $x=0$.

We start with the well known propagator $K_{ho}\left( x_f, t_f; x_i, t_i\right)$\footnote{See page 200 of Ref. \cite{Feynman}.} for a harmonic oscillator defined over the entire real line $(-\infty<x<\infty)$. It is
\begin{eqnarray}
K_{ho}\left( x_f, t_f; x_i, t_i\right) &\!\!\!\!=&\!\!\!\! \sqrt{\frac{m\omega}{2\pi\hbar\,i\,\sin\omega \tau}}~
\exp{\left(\frac{im\omega}{2\hbar\,\sin\omega \tau}\left[\left(x_f^2+x_i^2\right)\,\cos\omega \tau-2x_i\,x_f
\right]
\right)}
,~~
\end{eqnarray}
where we have substituted $\tau$ for $\left(t_f-t_i\right)$.

Then, assuming the applicability of the image method, the propagator for the H-HO is
\begin{eqnarray}
{\cal K}_{h-ho}\left( x_f, t_f; x_i, t_i\right)
&\!\!\!\!=&\!\!\!\! K_{ho}\left( x_f, t_f; x_i, t_i\right) - K_{ho}\left( x_f, t_f; -x_i, t_i\right)
\nonumber\\\nonumber\\
&\!\!\!\!=&\!\!\!\!
\sqrt{\frac{m\omega}{2\pi\hbar\,i\,\sin\omega \tau}}~
\exp{\left(\frac{im\omega}{2\hbar\,\sin\omega \tau}\left[\left(x_f^2+x_i^2\right)\,\cos\omega \tau
\right]
\right)}\times
\nonumber\\\nonumber\\&&
~~~~~~\left[
\exp{\left(-\,\frac{2\,im\omega\,x_i\,x_f}{2\hbar\,\sin\omega \tau}\right)}
-
\exp{\left(\frac{2\,im\omega\,x_i\,x_f}{2\hbar\,\sin\omega \tau}\right)}
\right]~~.\label{H-HO1}
\end{eqnarray}
We will now show that this expression generates the correct eigenvalues for the H-HO. We start with
\begin{eqnarray}
{\cal K}\left( x, t_f; x, t_i\right) =
\sum_{n=0}^\infty e^{-\frac{i}{\hbar}\,E_n \,\tau}\psi_n(x_f)\psi_n(x_i) \label{FKDecomp}~.
\end{eqnarray}
From here, we substitute $x_i = x_f = x$. This substitution and a subsequent integration  allows us to identify the eigenvalues:
\begin{eqnarray}
\int_0^\infty\dd x ~{\cal K}\left( x, t_f; x, t_i\right) &\!\!\!\!=&\!\!\!\!
\sum_{n=0}^\infty e^{-\frac{i}{\hbar}\,E_n \,\tau} \int_0^\infty\dd x ~\left|\psi_n(x)\right|^2
=
\sum_{n=0}^\infty e^{-\frac{i}{\hbar}\,E_n \,\tau}~,
\label{SpectralDecomposition}
\end{eqnarray}
where we have used the orthonormality of $\psi_n(x)$; i.e.,
$\int \dd x  \left|\psi_n(x)\right|^2 =1 $. Thus, the propagator ${\cal K}_{h-ho}\left( x, t_f; x, t_i\right)$ is then given by
\begin{eqnarray}
\!{\cal K}_{h-ho}\left( x, t_f; x, t_i\right) &\!\!\!\!=&\!\!\!\!
\sqrt{\frac{m\omega}{2\pi\hbar\,i\,\sin\omega \tau}}~\times ~~~~\label{H-HO2}
\left[
\exp{\left(\frac{im\omega x^2\,}{\hbar\,\sin\omega \tau}\left(\cos\omega \tau-1
\right)
\right)}
-
\exp{\left(\frac{im\omega x^2\,}{\hbar\,\sin\omega \tau}\left(\cos\omega \tau+1
\right)
\right)}
\right] ~.\nonumber
\end{eqnarray}


Then, as in Eq. (\ref{SpectralDecomposition}), we integrate the Gaussian ${\cal K}_{h-ho}\left( x, t_f; x, t_i\right)$
\begin{eqnarray}
\int_0^\infty\dd x ~{\cal K}_{h-ho}\left( x, t_f; x, t_i\right) &\!\!\!\!=&\!\!\!\!\frac{1}{2}\sqrt{\frac{m\omega}{2\pi\hbar\,i\,\sin\omega \tau}}~\times 
~~\left[
\sqrt{\frac{\pi i\,\hbar\,\sin\omega \tau}{m\omega\,\left(\cos\omega \tau-1
\right)}}
-
\sqrt{\frac{\pi i\,\hbar\,\sin\omega \tau}{m\omega\,\left(\cos\omega \tau-1
\right)}}~~
\right].
\end{eqnarray}
Further simplification leads to
\begin{eqnarray}
\int_0^\infty\dd x ~{\cal K}_{h-ho}\left( x, t_f; x, t_i\right) &\!\!\!\!=&\!\!\!\!\frac{1}{2} \left[
\frac{1}{\sqrt{2\left(\cos \omega \tau -1  \right)}}
-
\frac{1}{\sqrt{2\left(\cos \omega \tau +1  \right)}}
\right]
\end{eqnarray}
Now, using $1-\cos \omega \tau = 2 \sin^2\left(\frac{\omega \tau}{2}\right)$ and
$1+\cos \omega \tau = 2 \cos^2\left(\frac{\omega \tau}{2}\right)$, we get
\begin{eqnarray}
\!\!\!\!
\int_0^\infty\dd x ~{\cal K}_{h-ho}\left( x, t_f; x, t_i\right) &\!\!\!\!=&\!\!\!\!
\frac{1}{2} \left[
\frac{1}{2i \sin\frac{\omega \tau}2}
-
\frac{1}{2 \cos\frac{\omega \tau}2}
\right]=
\frac{1}{2}
\left[
\frac{1}{e^{\frac{i \omega \tau}2}-e^{-\frac{i \omega \tau}2}}
-
\frac{1}{e^{\frac{i\omega \tau}2}+e^{-\frac{i\omega \tau}2}}
\right]
\nonumber\\\nonumber\\&\!\!\!\!=&\!\!\!\!
\frac{1}{2}
\frac{2e^{-\frac{i\omega \tau}2}} {e^{i\omega \tau}-e^{-i\omega \tau} }
=
\frac{e^{-\frac{3i \omega \tau}2}} {1-e^{-2i\omega \tau} }
=e^{-\frac{3i\omega \tau}2}~\sum_{n=0}^\infty e^{-2i\,n \omega \tau}
=\sum_{n=0}^\infty
e^{-i\left( 2\, n + \frac32 \right) \omega \tau}~~. \label{SpectralDecomposition2}\nonumber
\end{eqnarray}
Thus, 
we see that we get the expected eigenvalues for the half-harmonic oscillator
\begin{eqnarray}
        E_n = \left( 2\, n + \frac32 \right) \hbar\omega\,, ~~n=0,1,2,3,\cdots .
\end{eqnarray}
Thus, the energy levels of the H-HO emerge to be the same as  the odd energy levels of the full  harmonic oscillator.

It is worth pointing out that our result could be arrived at much easier if we had started with
\begin{eqnarray}
K_{ho}\left( x_f, t_f; x_i, t_i\right) =
\sum_{n=0}^\infty e^{-\frac{i}{\hbar}\,\left(n+\frac12\right)\omega \,\tau} H_n\left(\sqrt{\frac{m\omega}{\hbar}}\, x_f\right)e^{-{\frac{m\omega}{2\hbar}}\,x_f^2}H_n\left(\sqrt{\frac{m\omega}{\hbar}}\,x_i\right)  e^{-{\frac{m\omega}{2\hbar}}\,x_i^2}
\end{eqnarray}
The propagator for the H-HO is then given by
\begin{eqnarray}
{\cal K}_{h-ho}\left( x_f, t_f; x_i, t_i\right) &\!\!\!\!=&\!\!\!\! K_{ho}\left( x_f, t_f; x_i, t_i\right) - K_{ho}\left( x_f, t_f; -x_i, t_i\right)
\nonumber\\\nonumber\\&\!\!\!\!=&\!\!\!\!
\sum_{n=0}^\infty e^{-i \left( n+\frac12\right) \,\omega \tau} e^{-{\frac{m\omega}{2\hbar}}\,\left(x_f^2+x_i^2\right)}
\left[
H_n\left(\sqrt{\frac{m\omega}{\hbar}}\, x_f\right) H_n\left(\sqrt{\frac{m\omega}{\hbar}}\,x_i\right)\right.
\nonumber\\\nonumber\\&&
\quad \quad \quad \quad
- \left.
H_n\left(\sqrt{\frac{m\omega}{\hbar}}\, x_f\right) H_n\left(-\sqrt{\frac{m\omega}{\hbar}}\,x_i\right)
\right]~.
\end{eqnarray}
Since $H_n(-x) = (-1)^n H_n(x)$, we find
\begin{eqnarray}
{\cal K}_{h-ho}\left( x_f, t_f; x_i, t_i\right) &\!\!\!\!=&\!\!\!\! \sum_{n=0}^\infty e^{-i \left( n+\frac12\right) \,\omega \tau} e^{-{\frac{m\omega}{2\hbar}}\,\left(x_f^2+x_i^2\right)}~
\left[1-(-1)^n  \right]
H_n\left(\sqrt{\frac{m\omega}{\hbar}}\, x_f\right) H_n\left(\sqrt{\frac{m\omega}{\hbar}}\,x_i\right)
\nonumber\\\nonumber\\ &\!\!\!\!=&\!\!\!\!
2\sum_{n~\rm odd}^\infty e^{-i \left( n+\frac12\right) \,\omega \tau} e^{-{\frac{m\omega}{2\hbar}}\,\left(x_f^2+x_i^2\right)}
H_n\left(\sqrt{\frac{m\omega}{\hbar}}\, x_f\right) H_n\left(\sqrt{\frac{m\omega}{\hbar}}\,x_i\right)
\nonumber\\\nonumber\\  &\!\!\!\!=&\!\!\!\!
\sum_{n=0}^\infty e^{-i \left( 2n+\frac32\right) \,\omega \tau}
\underbrace{\sqrt{2}\,e^{-{\frac{m\omega}{2\hbar}}\,x_f^2}
H_{2n+1}\left(\sqrt{\frac{m\omega}{\hbar}}\, x_f\right) }_{\phi_n(x_f)}
\underbrace{\sqrt{2}\,e^{-{\frac{m\omega}{2\hbar}}\,x_i^2} H_{2n+1}\left(\sqrt{\frac{m\omega}{\hbar}}\,x_i\right)}_{\phi_n(x_i)}.\nonumber\\
\end{eqnarray}
Note the factor of $\sqrt{2}$ in the normalization is necessitated by the change in domain. Thus, by comparing this result to Eq. (\ref{FKDecomp}), we see that the energy eigenvalues and corresponding eigenfunctions for the half-harmonic oscillator are given by
$$E_n = \left( 2\, n + \frac32 \right) \hbar\omega~~~~~~{\rm and}~~~~~~\phi_n(x) = \sqrt{2}\, e^{-\frac{m\omega}{2\hbar}\,x^2}
H_{2n+1}\left(\sqrt{\frac{m\omega}{\hbar}}\,x\right)$$
respectively.

Thus, beginning with two different initial points, we have shown that the image method readily generates the correct spectrum for the half-harmonic oscillator.

\vspace*{0.25in}
{\bf \Large Conclusion:}

We derived the propagator of a free particle in a restricted domain $(0<x<\infty)$ using eigenstates that vanish at the infinite boundary, showing that image contributions naturally emerge from this method. We also showed that the Rosen-Morse potential, for a specific value of a parameter, reduces to the Infinite Square Well potential, and that the corresponding Green's functions map into each other. Thus, starting from a well defined potential with no sharp boundary, we derived the Green's function for a system with two sharp boundaries and an infinite number of image charges. The power of our method lies in the fact that by employing a change of parameters there was no need for a point canonical transformation as in Ref. \cite{Sokmen-ISW}. Finally, we used the image method to derive the propagator for a half-harmonic oscillator and showed its correctness by generating a correct set of well known spectra for this system. This final case provides an explicit example of a particle in a non-zero potential with a sharp boundary and supports Aurbach and Schulman's proposition in Ref. \cite{Auerbach_Schulman}.


\newpage
\begin{thebibliography}{99}
\bibitem{Goodman} M. Goodman,
Am. Jour. Phys. {\bf 49}, 843 (1981).
%
\bibitem{Feynman} R. Feynman and A. Hibbs, Emended by Daniel F. Styer, ``{\it Quantum Mechanics and Path Integrals}", Dover Publications, Inc., Mineola, NY 2005.
\bibitem{Kleinert} H. Kleinert, ``{\it Path Integrals in Quantum Mechanics, Statistics, Polymer Physics, and Financial Markets}", 5th edition, World Scientific, Singapore, 2009.
\bibitem{Schulman}L. S. Schulman, ``{\it Techniques and Applications of Path Integration}", Dover Publications, Inc., Mineola, NY 2005.
%
\bibitem{Sokmen-ISW} I. S\"okmen,
Phys. Lett. {\bf 106 A}, 212 (1984).
%
\bibitem{Auerbach_Kivelson}A. Auerbach and S. Kivelson, Nucl. Phys. {\bf B 257},799 (1985).
%
\bibitem{Nevels_Wu_Huang}R. Nevels, Z. Wu, and C. Huang, Phys. Rev. {\bf A 48}, 3445 (1993).
%
\bibitem{Auerbach_Schulman} A. Auerbach and L. S. Schulman,
Jour. Phys A: Math Gen. {bf 30}, 5993 (1997).
%
\bibitem{Gervais} J. Gervais and A. Jevicki, Nucl. Phys. {\bf B 110}, 53, (1976).
%
\bibitem{Dutt2} 
    R. De, R. Dutt and U. P. Sukhatme, Phys.Rev. {\bf A 46}, 6869 (1992)
%
\bibitem{Dutt3}
R. De, R. Dutt and U. P. Sukhatme, Phys. Lett. {\bf A 191},352 (1994).
\bibitem{Abers_Lee}E. S. Abers and B. W. Lee, Phys. Report. {\bf 9\,C}, 1 (1973).
%
\bibitem{Sokmen-Rotator} N. K. Pak and I. S\"okmen,
Phys. Lett. {\bf 103 A}, 298 (1984).
%
\bibitem{Witten} E. Witten, Nucl. Phys. B202 (1982) 253.
\bibitem{Cooper1}F. Cooper and B. Freedman, Ann. Phys. 146 (1983) 262.
\bibitem{Junker} G. Junker, ``{\it \!Supersymmetric Methods in Quantum and Statistical Physics}", Springer-Verlag, Berlin, 1996.
\bibitem{Cooper-Khare-Sukhatme} F. Cooper, A. Khare and U. P. Sukhatme, ``{\it  \!Supersymmetry in Quantum Mechanics}", World Scientific, Singapore 2001.
\bibitem{Gangopadhyaya-Mallow-Rasinariu} For a recent review of SUSYQM see A. Gangopadhyaya, J. Mallow and C. Rasinariu, ``{\it  \!Supersymmetric Quantum Mechanics: An Introduction}", World Scientific, Singapore 2010.
%
\bibitem{Dutt} R. Dutt, A. Khare, and U. P. Sukhatme, Am. Jour. Phys. {\bf 56}, 163 (1988).
\bibitem{Hexagon}
A. Gangopadhyaya, 
P. K. Panigrahi, U. P. Sukhatme; Helvetica Physica
Acta. 67(4):363-368, (1994).
%
\bibitem{Infeld} L. Infeld and T. E. Hull {\it Rev. Mod. Phys} {\bf 23}, 21 (1951).
%
\bibitem{Gendenshtein} L. E. Gendenshtein {\it JETP Lett.} {\bf 38}, 356 (1983).

\end{thebibliography}
\end{document}